\documentclass[10pt]{article}
\usepackage{jheppub}  
\usepackage{hyperref}
\usepackage{graphicx,color}
  \usepackage{bm}
   \usepackage{amsmath}
    \usepackage{amssymb}
     \usepackage{pifont}
     \usepackage{multirow}

\newcommand{\nn}{\nonumber}

\newcommand{\ch}{\text{ch}}

\newcommand{\ot}{\leftarrow}

\renewcommand{\(}{\left(}
\renewcommand{\)}{\right)}
\renewcommand{\[}{\left[}
\renewcommand{\]}{\right]}

\renewcommand{\vec}[1]{\bm{#1}}

\newcommand{\specialcellcenter}[2][c]{\begin{tabular}[#1]{@{}c@{}}#2\end{tabular}}

\begin{document}
\title{Pion-induced Drell-Yan processes within TMD factorization}

\author{Alexey Vladimirov}
\affiliation{Institut f\"ur Theoretische Physik, \\ Universit\"at Regensburg,\\
D-93040 Regensburg, Germany}
\emailAdd{alexey.vladimirov@physik.uni-regensburg.de}

\abstract{
We extract the pion transverse momentum dependent (TMD) parton distribution by fitting the pion-induced Drell-Yan process within the framework of TMD factorization. The analysis is done at the next-to-next-to-leading order (NNLO) with proton TMD distribution and non-perturbative TMD evolution extracted earlier in the global fit. We observe the significant difference in the normalization of transverse momentum differential cross-section measured by E615 experiment and the theory prediction.
}
\maketitle

\section{Introduction}

Transverse momentum dependent (TMD) factorization theorem allows for a systematic study of partons transverse motions. Being equipped by the next-to-next-to-leading order (NNLO) evolution and matching, TMD factorization establishes an accurate framework for extractions of TMD distributions and production of trustful predictions for TMD cross-sections. It has been recently demonstrated in ref.\cite{Bertone:2019nxa} by the global analysis of Drell-Yan process. In this work, I extend the analysis of ref.\cite{Bertone:2019nxa} by considering the pion-induced Drell-Yan process and extracting the pion unpolarized TMD parton distribution function (TMDPDF). Apart of the pure scientific interest this study is stimulated by the upcoming measurement of the pion-induced Drell-Yan process at COMPASS facility \cite{Gautheron:2010wva}.

Formulated in \cite{Collins:1981va,Collins:1984kg}, TMD factorization theorem has been proven at all orders of perturbation theory \cite{Becher:2010tm,Collins:2011zzd,GarciaEchevarria:2011rb,Vladimirov:2017ksc}. Within the modern construct, the TMD distributions are generic non-perturbative functions that obey the double-scale evolution \cite{Vladimirov:2017ksc,Scimemi:2018xaf} and match collinear distributions at small-b limit \cite{Collins:2011zzd,GarciaEchevarria:2011rb,Becher:2011xn,Aybat:2011zv,Echevarria:2016scs,Gehrmann:2014yya,Scimemi:2019mlf,Scimemi:2019gge}. The matching to the perturbative limit almost guaranties the agreement with the high-energy data and the collinear factorization. Simultaneously, it greatly constraints the value of TMD distributions in the numerically dominant part of cross-section formula. As a result, the TMD factorized cross-section has a great predictive power even at low-energies, where influence of non-perturbative corrections is higher. Let me note, that NNLO perturbative input is important to describe precise modern data \cite{Scimemi:2017etj}. 

The TMD factorized cross-section contains three non-perturbative functions. There are two TMD distributions and the non-perturbative evolution kernel. It is practically difficult to decorrelate these functions. In ref.\cite{Bertone:2019nxa} the large data-set has been considered with significant difference in energy (from 4 to 150 GeV), which allowed to reduce the correlation between non-perturbative evolution and TMDPDFs. In this work, the situation is simpler since the non-perturbative evolution and the proton TMDPDF are taken from \cite{Bertone:2019nxa}. Therefore, the extraction pion TMDPDF is direct, and can be considered as a part of global fit of Drell-Yan data. 

The numerical part of the work has been done by \texttt{artemide} package \cite{web}. \texttt{Artemide} is the library of fotran modules related to different aspects of TMD factorization, from the small-b matching to the computation of the cross-section (including bin-integration and fiducial cuts, if required). The PDF sets are provided via the LHAPDF interface \cite{Buckley:2014ana}. The \texttt{artemide} repository also includes sets for TMD distributions (and their evolution) together with distributions of replicas. The results of the current extraction are added to the repository, as \texttt{Vpion19} set.

The pion-induced Drell-Yan process does not attract too much attention. For review of recent development see ref.\cite{Wang:2018syo}. Perhaps, the main reason is the low quality of the data. The last measurement has been done in the end of 80's at E615 at FermiLab \cite{Conway:1989fs}. In this work I have observed a systematic disagreement between E615-data and theory predictions in the normalization value. Currently, it is not possible to decide: is the disagreement a problem of theory or of the data. The similar problems have been observed recently in \cite{Bacchetta:2019tcu} in comparison of collinear factorization to low-energy Drell-Yan process. Hopefully, COMPASS results will resolve this issue.

The paper consists of two sections. The sec.\ref{sec:th}, I briefly review the TMD factorization framework, with the emphasis on difference between this work and ref.\cite{Bertone:2019nxa}, which consists in introduction of exact matching for $zeta$-line at large-$b$. The section \ref{sec:fit} is devoted to the comparison of the theoretical prediction to the data and to the extraction of the pion TMDPDF. The significant part of sec. \ref{sec:fit} is the discussion of the problem with the normalization for E615 measurement, and its possible origins. In the appendix \ref{app:app1} the derivation of the exact expression for the special null evolution line that was used for low-energy TMD evolution is presented.

\section{Theoretical framework}
\label{sec:th}

The derivation of the cross-section for Drell-Yan process in the TMD factorization has been a subject of many studies, see e.g. refs.\cite{Becher:2010tm,Collins:2011zzd,GarciaEchevarria:2011rb,Scimemi:2017etj}. In this section I present only the main formulas used in this analysis. The theory framework coincides with refs.\cite{Scimemi:2017etj,Bertone:2019nxa}. The particular points specific for discussed case are presented in details.

\paragraph{Cross-section within TMD factorization.}

The cross-section for $h_1+h_2\to \gamma^*(\to ll')+X$ is 
\begin{eqnarray}\label{th:xSec}
\frac{d\sigma}{dQ^2dx_Fdq_T^2}=\sigma_0\sum_{f_1,f_2}H_{f_1f_2}(Q,\mu)\int_0^\infty \frac{b d b}{2}J_0(b q_T)F_{f_1\ot h_1}(x_1,b;\mu,\zeta_1)F_{f_2\ot h_2}(x_2, b;\mu,\zeta_2),
\end{eqnarray}
where $q$ is the momentum of the photon, with the virtuality $q^2=Q^2$, and the transverse component $q_T$. Variable $x_F$ is the Feynman $x$ related to Bjorken $x$'s and $\tau$ in a usual manner,
\begin{eqnarray}
x_{1,2}=\frac{\pm x_F+\sqrt{x_F^2+4\tau}}{2},\qquad \tau=x_1 x_2= \frac{Q^2+\vec q_T^2}{s}.
\end{eqnarray}
The common factor for the cross-section is
\begin{eqnarray}\label{th:s0}
\sigma_0=\frac{4\pi \alpha_{\text{em}}^2(Q)}{9 Q^2 s \sqrt{x_F^2+4\tau}},
\end{eqnarray}
and the hard coefficient function is
\begin{eqnarray}\label{th:H}
 H_{f_1,f_2}(Q,\mu)=\sum_q \delta_{qf_1}\delta_{\bar q f_2}e_q^2\[1+2a_s(\mu)C_F\(-L^2+3L-8+\frac{7\pi^2}{6}\)+O(a_s^2)\],
\end{eqnarray}
where sum over $q$ runs though quarks and anti-quarks, $L=\ln(Q^2/\mu^2)$, $C_F=4/3$. The NNLO term of the hard coefficient function used in current evaluation can be found in \cite{Gehrmann:2010ue}. The functions $F_{f\ot h}(x,b;\mu,\zeta)$ in (\ref{th:xSec}) are TMDPDF for parton $f$ in the hadron $h$ evaluated at the scale $(\mu,\zeta)$.

\paragraph{Selection of scales and TMD evolution.}

The scales $\mu^2$, $\zeta_1$ and $\zeta_2$ are of order of $Q^2$. To be specific, we fix $\mu=Q$, so $L=0$ in the expression for hard coefficient function (\ref{th:H}), and $\zeta_1=\zeta_2=Q^2$, so $\zeta_1\zeta_2=Q^4$ as it is defined within TMD factorization \cite{Collins:2011zzd,Vladimirov:2017ksc,GarciaEchevarria:2011rb}. From the hard scale $(\mu,\zeta)=(Q,Q^2)$ TMDPDFs are evolved to the defining scale with the help of the TMD evolution, \cite{Aybat:2011zv,Collins:2011zzd,Scimemi:2018xaf}. The defining scale for TMDPDF is selected in accordance to $\zeta$-prescription \cite{Scimemi:2017etj,Scimemi:2018xaf}. 

In $\zeta$-prescription, TMDPDFs are defined at the line $\zeta=\zeta(\mu,b)$, which is a null-evolution line in the plane $(\mu,\zeta)$. The optimal TMD distribution used in this work, belongs to the null-evolution line that passes through the saddle point of the evolution field. This boundary condition is very important for two reasons. First, there is only one saddle point in the TMD evolution field, and thus the special null-evolution line is unique. Second, the special null-evolution line is the only null-evolution line, which has finite $\zeta$ at all values of $\mu$ (with $\mu$ bigger than $\Lambda_{QCD}$). It follows from the definition of the saddle point, and guaranties the finiteness of perturbative series at each order. The optimal distribution is denoted as $F_{f\to h}(x,b)$ (without scale arguments), what emphasizes its uniqueness and ``naive'' scale-invariance. The relation between the optimal TMD distribution and TMD distribution at the scale $(\mu,\zeta)=(Q,Q^2)$ is
\begin{eqnarray}
F_{f\ot h}(x,b;Q,Q^2)=\(\frac{Q^2}{\zeta_{\text{NP}}(Q,b)}\)^{-\mathcal{D}_{\text{NP}}(b,Q)}F_{f\ot h}(x,b),
\end{eqnarray}
where $\mathcal{D}$ is the rapidity anomalous dimension. The derivation of this simple expression and proof of its equivalence to the standard Sudakov exponent is given in \cite{Scimemi:2018xaf}. The subscript NP on the rapidity anomalous dimension $\mathcal{D}_{\text{NP}}$ and the special null-evolution line $\zeta_{\text{NP}}$ stresses the presence of non-perturbative corrections in both objects.

\paragraph{Expression for TMDPDF.} 

There are two places where non-perturbative physics enters TMD factorized cross-section. The first is the TMDPDFs $F(x,b)$ that describes transverse motion of confined quarks in a hadron. The second is the rapidity anomalous dimension $\mathcal{D}_{\text{NP}}(\mu,b)$ that describes the long-range correlation of gluons in QCD vacuum. So, non-perturbative structure of these objects are related to different aspects of QCD dynamics and are completely independent. At small values of $b$ both $F(x,b)$ and $\mathcal{D}_{\text{NP}}(\mu,b)$ could be calculated by means of the operator product expansion, see e.g. \cite{Aybat:2011zv,Bacchetta:2013pqa,Echevarria:2016scs,Scimemi:2019mlf,Scimemi:2019gge}. At large-$b$ the values of these function should be calculated in non-perturbative models, as e.g. in refs.\cite{Schweitzer:2012hh,Lorce:2014hxa,Noguera:2015iia}, or extracted from the data, as e.g. in refs.\cite{Su:2014wpa,DAlesio:2014mrz,Bacchetta:2017gcc,Scimemi:2017etj,Bertone:2019nxa}. 

The convenient ansatz that merges perturbative and non-perturbative part of TMDPDF functions is
\begin{eqnarray}\label{th:TMDPDF}
F_{f\ot h}(x,b)=\sum_{f'}\int_x^1 \frac{dy}{y}C_{f\ot f'}(y,b,\mu)f_{1,f'\ot h}\(\frac{x}{y},\mu\)f_{\text{NP}}(x,b),
\end{eqnarray}
where $C$ is the perturbative coefficient function calculated at NNLO in \cite{Gehrmann:2014yya,Echevarria:2016scs}, $f_1$ is unpolarized collinear PDF, and $f_{\text{NP}}$ is a non-perturbative modification function. The function $f_1$ must turn to $1$ at $b\to 0$. The selection of an ansatz for $f_{\text{NP}}$ is a delicate process, since it is the main source of biases, for a more detailed discussion see sec.2 in ref.\cite{Bertone:2019nxa}.

In the present work, the proton TMDPDF is taken from ref. \cite{Bertone:2019nxa}, where it was extracted from the global fit of high-energy (Tevatron and LHC) and low-energy (FermiLab and PHENIX) Drell-Yan measurements. The analyzed measurements (E537 \cite{Anassontzis:1987hk}, E615 \cite{Conway:1989fs} and NA3 \cite{Badier:1982zb}) were made on a tungsten(E537,E615) and platinum(NA3) targets ($Z=74$, $A=184$ and $Z=78$, $A=195$). Therefore, the proton TMDPDF from \cite{Bertone:2019nxa} requires a modification to simulate the nuclear environment. It is done by the rotation of the iso-spin components only. For example, for u-quark the nuclear TMDPDF is
\begin{eqnarray}\label{th:iso-spin-rotation}
F_{u\ot A}(x,b)=\frac{Z}{A}F_{u\ot p}(x)+\frac{A-Z}{A}F_{u\ot n}(x)=\frac{Z}{A}F_{u\ot p}(x)+\frac{A-Z}{A}F_{d\ot p}(x),
\end{eqnarray}
and similar for $d$, $\bar u$ and $\bar d$ distributions. 

The values of pion TMDPDF are fit to the data, as discussed in the following. The collinear pion PDF is taken \texttt{JAM18pionPDFnlo}-set \cite{Barry:2018ort}. The function $f_{\text{NP}}$ is taken similar to those used for proton in \cite{Bertone:2019nxa}. Taking into account the the fact that typical values of $x$ in the pion-induced Drell-Yan process are very high, the terms relevant for low-$x$ values were dropped. The resulting function depends on three parameters and reads
\begin{eqnarray}\label{th:ansatz}
f_{\text{NP}}^\pi(x,b)=\exp\(-\frac{(a_1+(1-x)^2 a_2)b^2}{\sqrt{1+a_3b^2}}\).
\end{eqnarray}
The parameters $a_{1,2,3}$ are to be fit to the data. Generally speaking, the non-perturbative function $f_{NP}$ depends on the flavor of parton. This dependence is ignored here, since the quality of analyzed data does not allow a flavor separation.

\paragraph{Expression for $\mathcal{D}_{\text{NP}}$.} 

The non-perturbative expression for $\mathcal{D}_{\text{NP}}$ has been extracted in \cite{Bertone:2019nxa} together with proton TMDPDF. It has the following form
\begin{eqnarray}\label{th:DNP}
\mathcal{D}_{\text{NP}}(\mu,b)=\mathcal{D}_{\text{pert}}(\mu,b^*(b))+d_{\text{NP}}(b),
\end{eqnarray}
where $\mathcal{D}_{\text{pert}}(\mu,b)$ is the perturbative part of rapidity anomalous dimension calculated at NNLO in \cite{Becher:2010tm,Echevarria:2015byo}, and at N$^3$LO in \cite{Vladimirov:2016dll,Li:2016ctv}. The function $d_{\text{NP}}(b)$ is the non-perturbative correction. In (\ref{th:DNP}) the resummed version of $\mathcal{D}_{\text{pert}}(\mu,b)$ \cite{Echevarria:2012pw,Scimemi:2018xaf} is used. The resummed expression for $\mathcal{D}_{\text{pert}}$ contains Landau pole at large values of $b$. To avoid it, the parameter $b$ is replaced by $b^*(b)$ in (\ref{th:DNP}),
\begin{eqnarray}
b^*(b)=b\Big/\sqrt{1+\frac{b^2}{B_{\text{NP}}^2}}.
\end{eqnarray}
The non-perturbative function $d_{\text{NP}}$ is
\begin{eqnarray}
d_{\text{NP}}(b)=c_0 b b^*(b).
\end{eqnarray}
The parameters $B_{\text{NP}}$ and $c_0$ are fit in \cite{Bertone:2019nxa}, and reads $B_{\text{NP}}=2.29\pm 0.43$, $c_0=0.022\pm 0.009$.

The only difference in the theory implementation between this work and ref.\cite{Bertone:2019nxa} is the expression for $\zeta_{\text{NP}}$. In ref.\cite{Bertone:2019nxa}, $\zeta_{\text{NP}}$ was modeled $\zeta_{\text{NP}}(b)=\zeta_{\text{pert}}(b^*(b))$. It corresponds to the special null-evolution line derived for $\mathcal{D}_{\text{NP}}=\mathcal{D}_{\text{pert}}(\mu,b^*(b))$, ignoring the $d_{\text{NP}}$-contribution. This choice is almost perfect for $d_{NP}\ll \mathcal{D}_{\text{pert}}$,  but this model deviates from the exact $\zeta_{\text{NP}}$ significantly for larger $d_{\text{NP}}$ (that happens at $b>B_{\text{NP}}$). The exact $\zeta_{\text{NP}}$ is $\zeta_{\text{NP}}$ determined by its differential equation (\ref{th:sp-line}). The deviation of $\zeta_{\text{NP}}$ from its exact values at large-$b$ could be seen as a part of non-perturbative model. However, it adds an undesired correlation between TMDPDFs and $\mathcal{D}_{\text{NP}}$ at large-$b$. The only reason to use the model values for $\zeta_{\text{NP}}$ in ref.\cite{Bertone:2019nxa} was the absence of a way to find exact $\zeta_{\text{NP}}$ at large $b$, where the saddle point runs to $\mu<\Lambda_{QCD}$ region. This problem has been solved recently by using the $\mathcal{D}_{\text{NP}}$ as an independent variable that accumulates all non-perturbative information and $b$-dependence. In this case, expression for exact $\zeta_{\text{NP}}$ can be found order-by-order in $a_s$ which is the only parameter. The details of calculation and explicit expression for $\zeta_{\text{exact}}$ are given in appendix \ref{app:app1}.

One of the requirement of the small-b matching procedure (\ref{th:TMDPDF}) in the $\zeta$-prescription is that values of $\zeta$-line should exactly match its pure perturbative  expression at $b\to 0$. Otherwise the exact cancellation of divergent $\ln(b^2)$ in the matching coefficient $C(x,b,\mu)$ in (\ref{th:TMDPDF})  does not take place \cite{Scimemi:2018xaf}. In order to facilitate the cancellation, the following form for $\zeta_{\text{NP}}$ has been used
\begin{eqnarray}\label{th:zetaNP}
\zeta_{\text{NP}}(\mu,b)=\zeta_{\text{pert}}(\mu,b)e^{-\vec b^2/B_{\text{NP}}^2}+\zeta_{\text{exact}}(\mu,b)\(1-e^{-\vec b^2/B_{\text{NP}}^2}\).
\end{eqnarray}
This form $\zeta_{\text{NP}}$ exactly matches $\zeta_{\text{pert}}$ at $b\ll B_{\text{NP}}$ and smoothly turns to exact value.

\paragraph{Perturbative orders.} Let me summarize the orders of perturbation theory are used in this work: 
\begin{itemize}
\item Hard coefficient function $H_{ff'}(\mu,Q)$ in (\ref{th:xSec}) is taken at NNLO (i.e. up to $a_s^2$-terms inclusively) \cite{Gehrmann:2010ue}.
\item Matching coefficient for unpolarized TMDPDF $C_{f\to f'}(x,b)$ in (\ref{th:TMDPDF}) is taken at NNLO (i.e. up to $a_s^2$-terms inclusively)  \cite{Echevarria:2016scs}, in $\zeta$-prescription \cite{Scimemi:2017etj}.
\item The perturbative part of rapidity anomalous dimension $\mathcal{D}_{\text{perp}}(\mu, b)$ in (\ref{th:DNP}) is taken at NNLO (i.e. up to $a_s^2$-terms inclusively) \cite{Echevarria:2015byo}, in the resummed form \cite{Echevarria:2012pw,Scimemi:2018xaf}.
\item The $\zeta_{\text{pert}}(\mu,b)$ in (\ref{th:zetaNP}) is taken at NNLO (i.e. up to $a_s^2$-terms inclusively) \cite{Scimemi:2018xaf}.
\item The $\zeta_{\text{exact}}(\mu,b)$ in (\ref{th:zetaNP}) is taken at NNLO (i.e. up to $a_s^1$-terms inclusively), see eqns.(\ref{app:g}-\ref{app:g2}).
\item To evaluate expressions for last three points one needs cusp anomalous dimension and $\gamma_V$ anomalous dimension up to  $a_s^3$-terms and $a_s^2$-terms, receptively. They could be found in \cite{Moch:2004pa,Moch:2005tm}.
\end{itemize}
Thus, the computation is done at complete NNLO perturbative accuracy.

\section{Comparison to the data}
\label{sec:fit}
\begin{table}[b]
\begin{center}
{\small
\begin{tabular}{|l|c|c|c|c|c|c|}
Experiment & $\sqrt{s}$[GeV] & $Q$[GeV] & $x_F$ & $N_{pt}$ & corr.err. & \specialcellcenter{Typical\\ stat.err.}
\\\hline\hline
E537 ($Q$-diff.) & 15.3 & \specialcellcenter{$4.0<Q<9.0$\\ in 10 bins} & $-0.1<x_F<1.0$ & 60/146 & 8\% & $\sim 20\%$
\\\hline
E537 ($x_F$-diff.) &  15.3 & $4.0<Q<9.0$ & \specialcellcenter{$-0.1<x_F<1.0$\\ in 11 bins} & 110/165 & 8\% & $\sim 20\%$
\\\hline\hline
E615 ($Q$-diff.) & 21.8 & \specialcellcenter{$4.05<Q<13.05$\\ in 10(8) bins} & $0.0<x_F<1.0$ & 51/155 & 16\% & $\sim 5\%$
\\\hline
E615 ($x_F$-diff.) & 21.8 & $4.05<Q<8.55$ & \specialcellcenter{$0.0<x_F<1.0$\\ in 10 bins} & 90/159 & 16\% & $\sim 5\%$
\\\hline\hline
\multirow{2}{*}{NA3} & \multirow{2}{*}{\specialcellcenter{16.8 , 19.4 \\ 22.9}} & $4.1<Q<8.5$ & $y>0$(?) & \multirow{2}{*}{--} & \multirow{2}{*}{15\%} & \multirow{2}{*}{--}
\\\cline{3-4}
 &  & $4.1<Q<4.7$ & $0<y<0.4$ &  &  & 
\\\hline
\end{tabular}}
\end{center}
\caption{\label{tab:data} The synopsis on the data used in the work. $N_{pt}$ is the number of points in the data set after/before the application of TMD factorization cut. Typical statistical error is estimated from the first 3 points for each $(Q,x_F)$-bin, and presented only for demonstration purposes. The data for NA3 is available only as a figure (figs. 1 and 2 in ref.\cite{Badier:1982zb}).}

\end{table}

\paragraph{Review of available data.}
There are three available measurements of transverse momentum cross-section for pion-induced Drell-Yan process. They were performed by NA3 \cite{Badier:1982zb}, E537 \cite{Anassontzis:1987hk}  and E615 \cite{Conway:1989fs} experiments. The measurement by NA3 is presented only by a plot in ref.\cite{Badier:1982zb}, and  the exact values of data-points and their error-bars are not available. Therefore, only the visual comparison with NA3 is possible (fig.\ref{fig:NA3}). The data tables for E537 \cite{Anassontzis:1987hk} and E615 \cite{Conway:1989fs} can be can be found in \cite{Stirling:1993gc}.

Both experiments E537 and E615 have been performed in the same environment at different energies of the pion beam, $P_\text{beam}=125$GeV for E537 and $P_\text{beam}=252$ GeV for E615, which corresponds to $s=235.4$GeV$^2$ and $s=473.6$GeV$^2$, respectively. The data for both experiments are provided in two alternative binning: differential in $x_F$, or differential in $Q$. In table \ref{tab:data}, the summary of kinematics for each data set is shown. Let me mention that both of measurements are made at high values of $x_{1,2}$. In particular, the lowest accessible value of $x_\pi$ is $0.26$ (for E537) and $0.18$ (for E615).

\paragraph{Definition of $\chi^2$ distribution.} 
To estimate the theory-to-data agreement, I have used the $\chi^2$-test function contracted as usual
\begin{eqnarray}
\chi^2=\sum_{ij}(m_i-t_i)V_{ij}^{-1}(m_j-t_j),
\end{eqnarray}
where $m_i$ is the central value of $i$'th data-point, $t_i$ is the theory prediction for this data-point and $i,j$ run through all points in the set. Both experiments provide an uncorrelated error for each point $\sigma_{\text{stat}}$ and a systematic error. The later is mainly generated by the luminosity uncertainty, and thus can be considered as a correlated error $\sigma_{\text{corr}}$. The covariance matrix is be build according to general rules:
\begin{eqnarray}\label{ex:V}
V_{ij}=\sigma_{\text{stat},i}^2\delta_{ij}+\sigma_{\text{corr},i}\sigma_{\text{corr},j}.
\end{eqnarray}

As it is discussed below, E615 data-set has a problem with the general normalization (it could be also a theory problem). Due to it, the value of $\chi^2$ calculated with covariance matrix (\ref{ex:V}) is extremely high, despite the errors of measurements are relatively large. It happens because the correlated part of $\chi^2$ overweights the uncorrelated part by an order of magnitude. So, the fit procedure becomes impossible. To stabilize the values of $\chi^2$, I have split the data-set of E615 to subsets with the same values of $(Q,x_F)$. Consequently, the correlated error has been adjusted to $(Q,x_F)$-bin independently. In other words, the elements of covariance matrix that mixes different $(Q,x_F)$-bins are set to zero. The possible sources of this problem are discussed below.

\begin{figure}[t]
\includegraphics[width=0.33\textwidth]{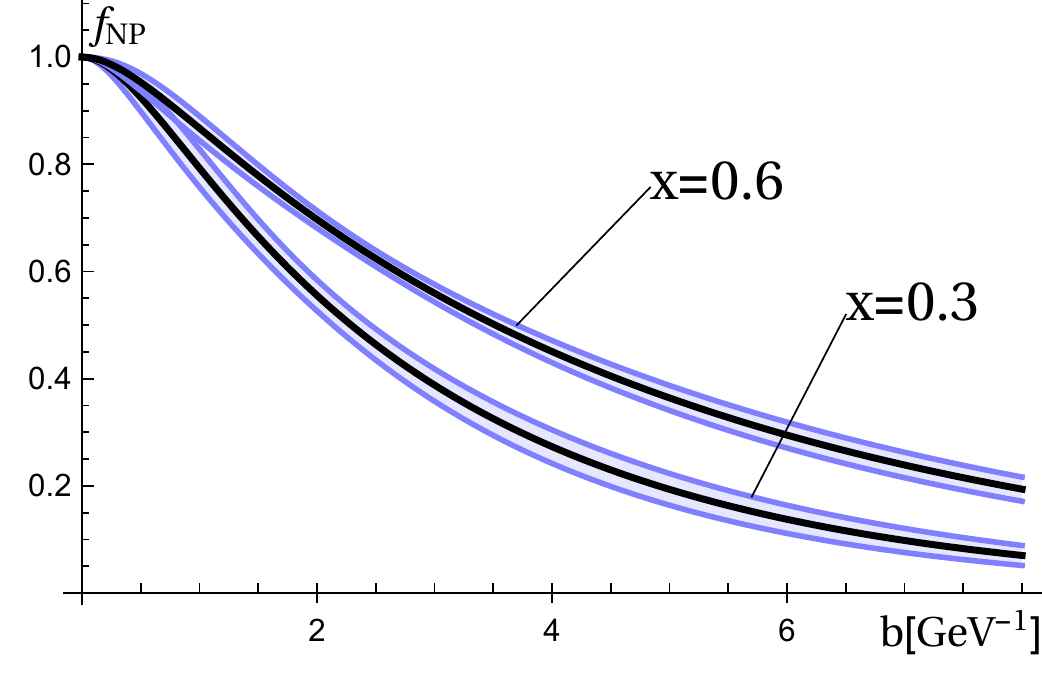}
\includegraphics[width=0.33\textwidth]{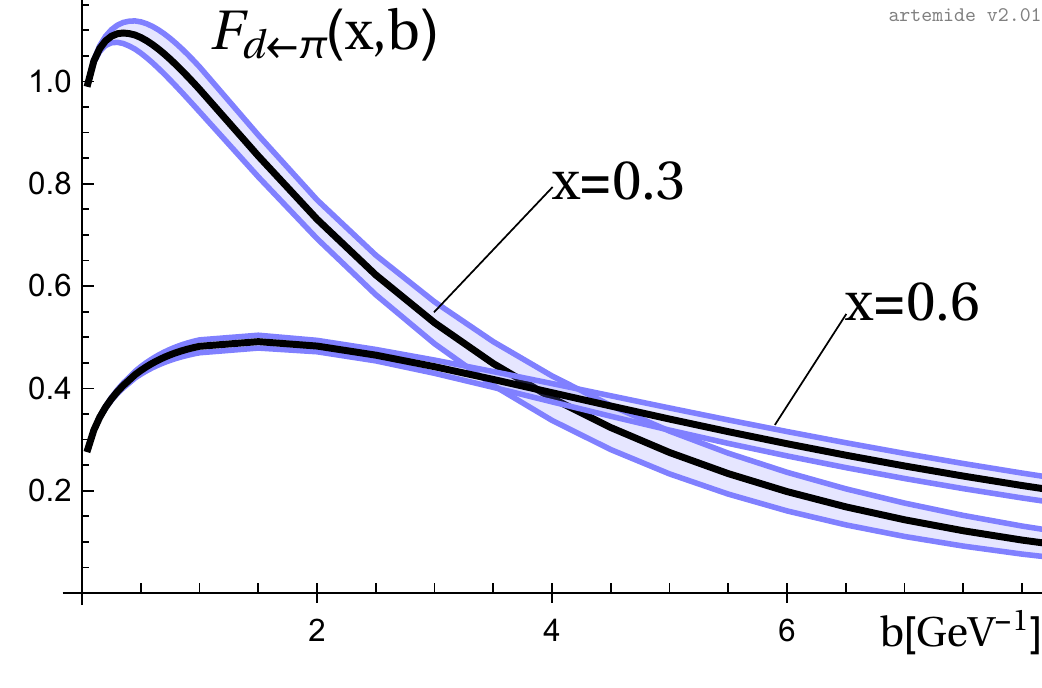}
\includegraphics[width=0.33\textwidth]{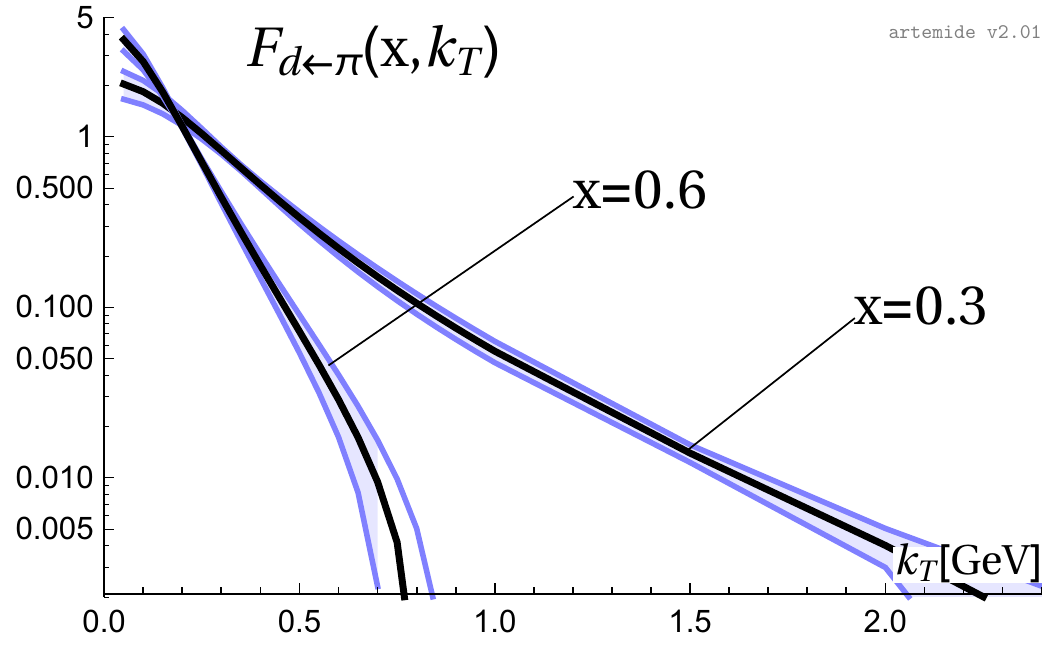}
\caption{\label{fig:fNP} (left) The function $f_{NP}$ that parametrizes the non-perturbative part of TMDPDF for pion \ref{th:TMDPDF}. (center) Pion TMDPDF for $d$-quark in $b$-space. (right) Pion TMDPDF for $d$-quark in $k_T$-space. The bands are the 1$\sigma$ uncertainty band related to the data error-bands and calculated by the replica method.}
\end{figure}

\paragraph{Selection of the data for the fit.}
The TMD factorization formula is derived in assumption that $q_T/Q$ is small. Practically, it is realized by considering the points with $q_T<\delta\cdot Q$, where $\delta\approx 0.25$ as it has been derived in \cite{Scimemi:2017etj} from the global analysis of Drell-Yan measurements. For $x_F$-differential measurements that have wide Q-bins, the center of Q-bin is used, what corresponds to $q_T\lesssim 2.2$GeV. 

There are 4 data-sets listed in table \ref{tab:data}. Data-sets belonging to the same experiment could not be added to a single $\chi^2$, since it would imply a double counting of a measurement. Doubtless, it is preferable to consider $x_F$-differential bins, since the $Q$-dependence is dictated by the evolution that is fixed from other data. Therefore, \textit{for the fit of non-perturbative parameters only the E615 differential in $x_F$ data-set has been used.} Furthermore, the bins $x_F\in(0.8,0.9)$ and $x_F\in(0.9,1.0)$ have been excluded, because there $x_\pi\sim 1$ and thus, the threshold resummation must be applied. The resulting set has 80 points.

\begin{figure}[t]
\includegraphics[width=\textwidth]{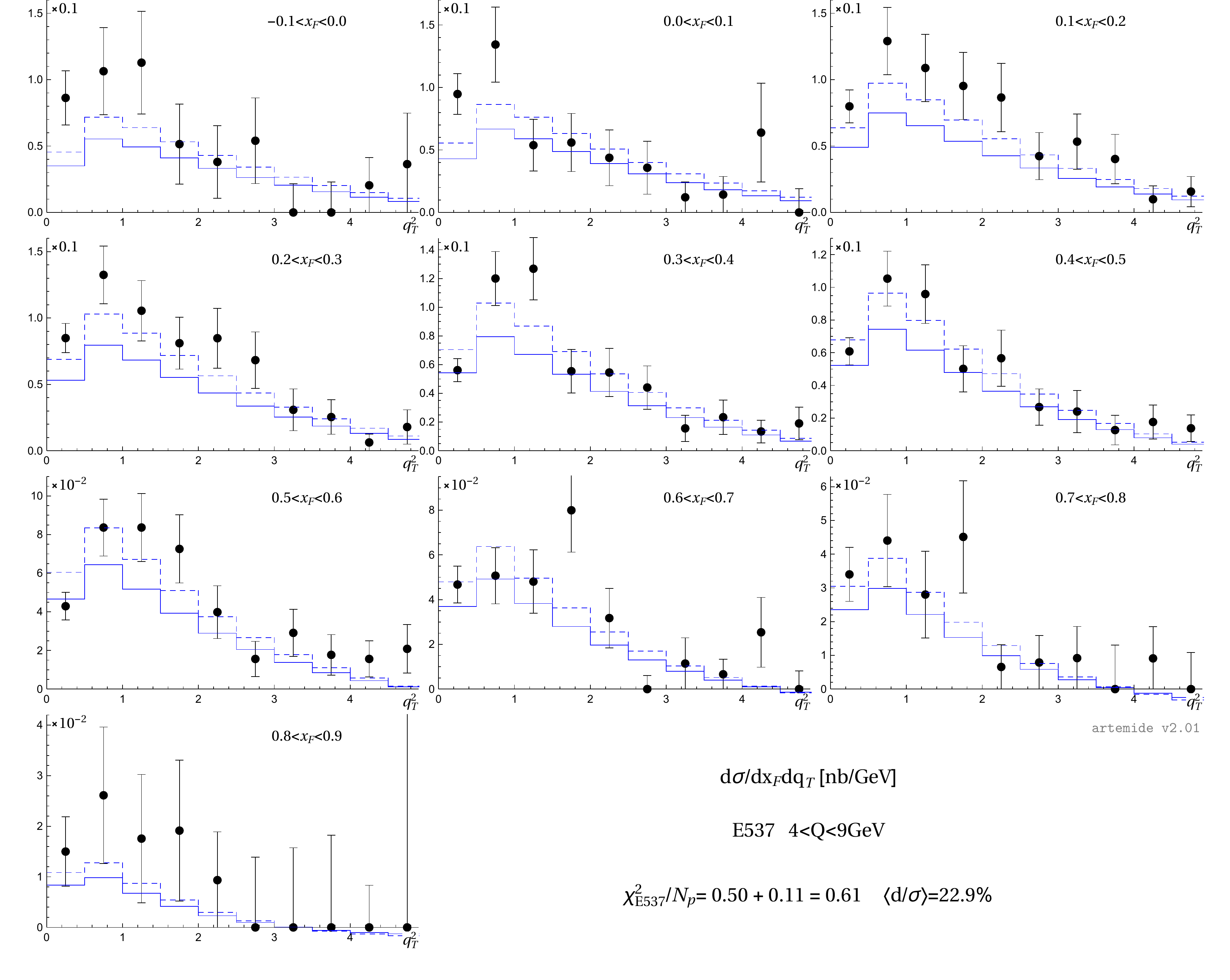}
\caption{\label{fig:E537-dX} Comparison of the theory prediction (solid line) to E537 differential in $x_F$. The dashed line is the theoretical prediction after the addition of systematic shifts $d_i$. The values of the $\chi^2$ and $d_i$ are calculated for the full set of data with 8\% correlated error.}
\end{figure}

\begin{figure}[t]
\includegraphics[width=\textwidth]{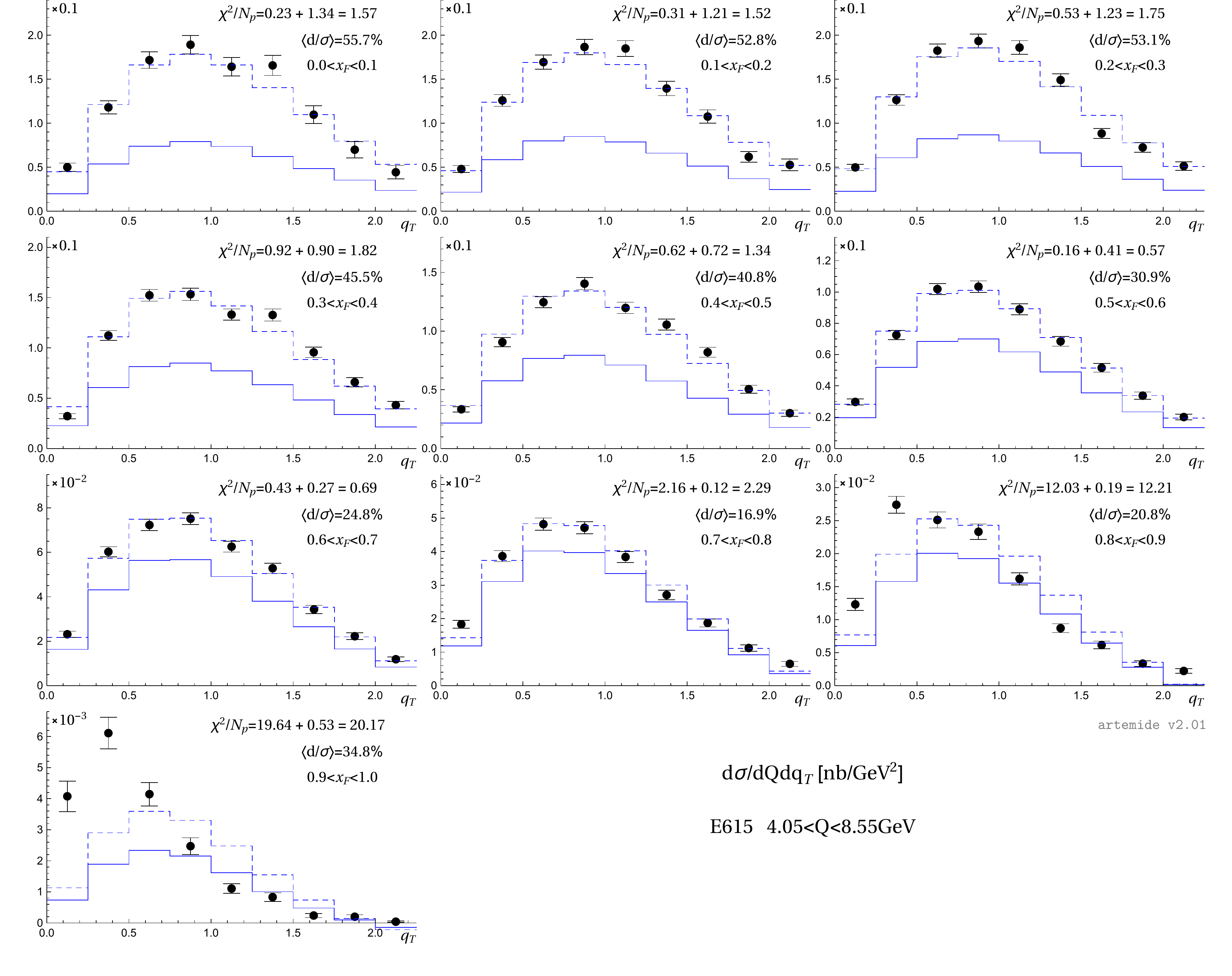}
\caption{\label{fig:E615-dX} Comparison of the theory prediction (solid line) to E615 differential in $x_F$. The dashed line is the theoretical prediction after the addition of systematic shifts $d_i$. The values of the $\chi^2$ and $d_i$ are calculated for the each $x_F$-bin with 16\% correlated error.}
\end{figure}

\begin{figure}[t]
\includegraphics[width=\textwidth]{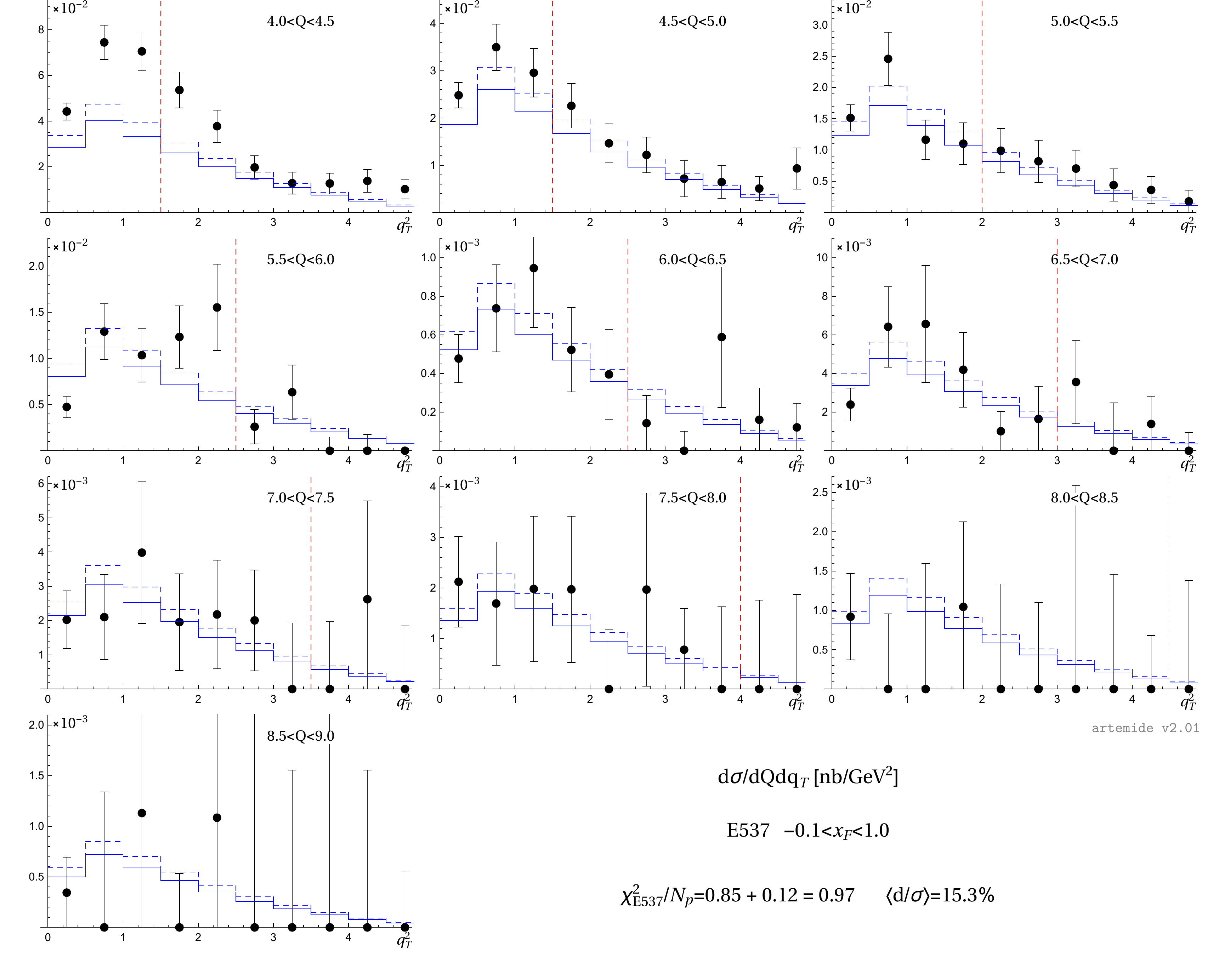}
\caption{\label{fig:E537-dQ} Comparison of the theory prediction (solid line) to E537 differential in $Q$. The dashed line is the theoretical prediction after the addition of systematic shifts $d_i$. The values of the $\chi^2$ and $d_i$ are calculated for the full set of data with 8\% correlated error. The vertical dashed line shows the estimation of the boundary for TMD factorization approach.}
\end{figure}

\begin{figure}[t]
\includegraphics[width=\textwidth]{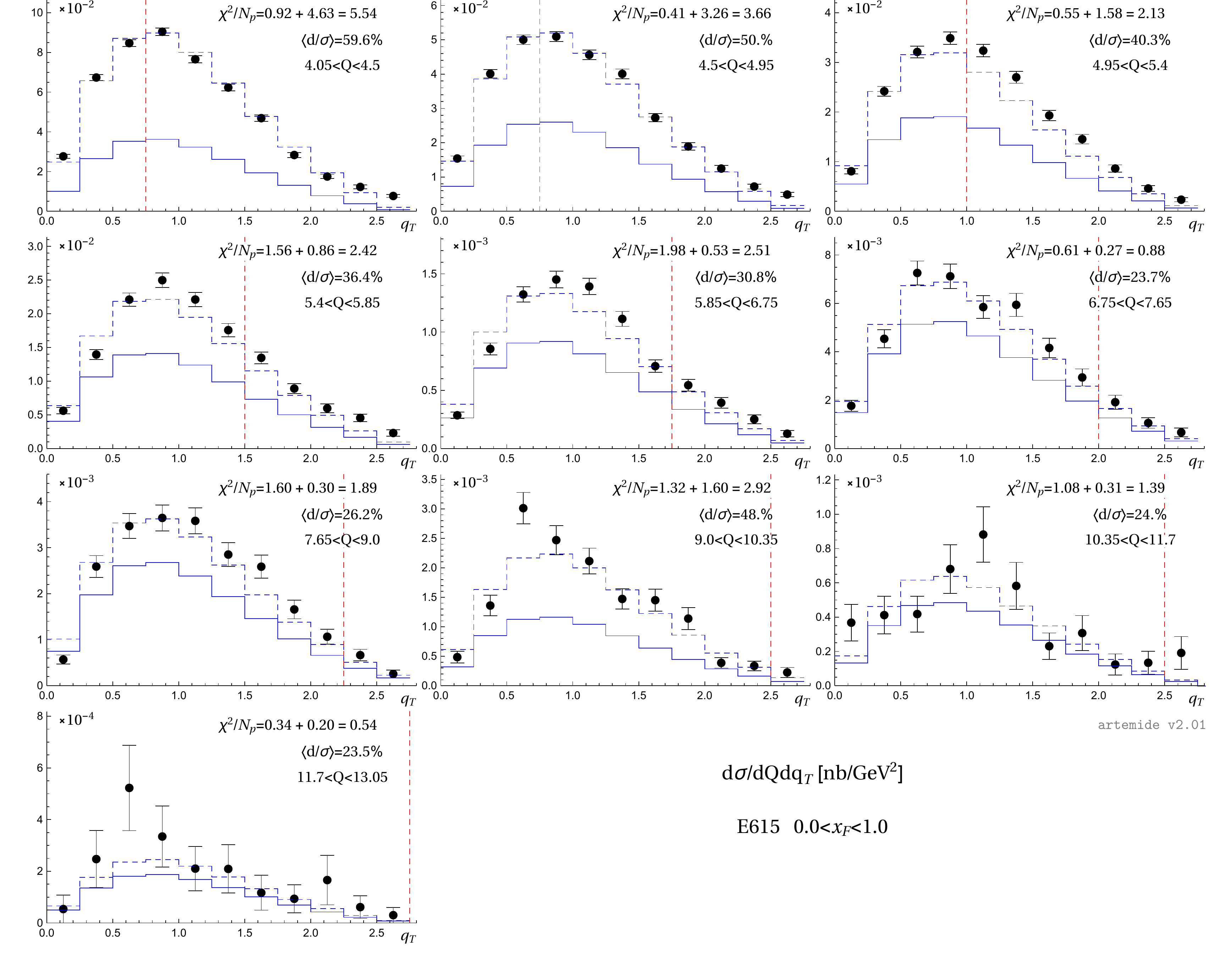}
\caption{\label{fig:E615-dQ} Comparison of the theory prediction (solid line) to E615 differential in $Q$. The dashed line is the theoretical prediction after the addition of systematic shifts $d_i$. The values of the $\chi^2$ and $d_i$ are calculated for the each $Q$-bin with 16\% correlated error.  The vertical dashed line shows the estimation of the boundary for TMD factorization approach. Note, that the bins with $Q\in(9,10.35)$ and $Q\in (10.35,11.7)$ lies in the region of $\Upsilon$-resonance, and could not be described by pure perturbative approach.}
\end{figure}

\paragraph{Dependence on collinear PDFs.} 
According to (\ref{th:TMDPDF}), the values of TMDPDF depend on collinear PDF. The dependence is partially compensated by the non-perturbative parameters of TMDPDF, that are fit separately for each PDF set. Nonetheless, the values of TMDPDF based on different PDF sets could significantly vary. The choice of PDF set also affects the non-perturbative TMD evolution, although in a lesser amount. The original \texttt{BSV19} extraction uses \texttt{NNPDF3.1} set of collinear PDFs \cite{Ball:2017nwa}. Additionally, the extraction of TMDPDFs and $\mathcal{D}_{\text{NP}}$ based on different collinear PDFs were performed (the analysis of these results will be presented elsewhere \cite{TOBE}), and they are available at \cite{web}. 

In the present study, I have compared the predictions generated with proton TMDPDFs (and $\mathcal{D}_{\text{NP}}$) based on different collinear PDF, and found results alike.  Particularly, $\chi^2$-minimization with proton TMDPDFs based on \texttt{MMHT14}(nnlo) \cite{Harland-Lang:2014zoa}, \texttt{NNPDF3.1} (nnlo) \cite{Ball:2017nwa} and \texttt{HERA20PDF} (nnlo) \cite{Abramowicz:2015mha} gives $\chi^2/N_p=1.45$, $1.70$ and $1.44$, correspondingly.  Taking into account, that \texttt{HERA20PDF} set also shows better global $\chi^2$ on the data-set from ref.\cite{Bertone:2019nxa}, in the following  \textit{the proton TMDPDF and non-perturbative part of TMD evolution is based on HERA20PDF are used}. This set \texttt{BSV19.HERA20PDF} can be downloaded from \texttt{artemide} repository \cite{web}. For pion collinear PDF  \texttt{JAM18pionPDF}-set has been used \cite{Barry:2018ort}.

\paragraph{Results of the fit.}
The minimization procedure for $\chi^2$-test yields the following values of non-perturbative parameters
\begin{eqnarray}\label{ph:a123}
a_1=0.17\pm 0.11\pm 0.03,\qquad a_2=0.48\pm 0.34\pm 0.06,\qquad a_3=2.15\pm 3.25\pm 0.32.
\end{eqnarray}
The first error-band is due to the uncertainty of data-points. It is estimated by the replica method, as in ref.\cite{Ball:2008by}, by minimization of $\chi^2$ on 100 replicas of pseudodata. The second error is due to uncertainty in the proton TMDPDF and TMD evolution. It is estimated by the minimization of $\chi^2$ on 100 of replicas of input distributions. 

Parameters $a_{1,2,3}$ are restricted to positive values. So, large error-bands in (\ref{ph:a123}) are the result of very asymmetric distribution of parameters. Large error bands on parameters does not implies a significant point-by-point uncertainty for $f_{\text{NP}}$, since all parameters are correlated. For example, at $b\sim 0.5$GeV$^{-1}$ the uncertainty in $f_{\text{NP}}$ is $2-3\%$. However, this band is definitely biased by the ansatz (\ref{th:ansatz}). The plot for $f_{\text{NP}}$ is shown in fig.\ref{fig:fNP}(left). The actual values of TMDPDF in $b-$space and $k_T$-space (that is obtained by Fourier transformation) are shown in fig.\ref{fig:fNP}(center,right). The pion TMDPDF obtained in this work together with distribution of 100 replicas is available in the \texttt{artemide}-repository \cite{web} as \texttt{Vpion19} TMDPDF set (for $\pi^-$-meson).

The final values of $\chi^2$ is $\chi^2/N_p=1.44$ $(N_p=80)$. It can be compared with the result of fit in ref.\cite{Wang:2017zym} $\chi^2/N_p=1.64$, where almost the same data were used. The main contribution to the value of $\chi^2$ comes from the systematic disagreement in the normalization between the data and the theory. In fig.\ref{fig:E537-dX},\ref{fig:E615-dX},\ref{fig:E537-dQ},\ref{fig:E615-dQ} the comparison of the data to the theory prediction is shown together with the values of $\chi^2/N_p$ for a given subset of data-points. In fig.\ref{fig:NA3} the visual comparison of the theory to NA3 is shown. The plots for $Q$-differential bins made for the range of $q_T$ larger than it is allowed by the TMD factorization (the boundary $q_T \simeq 0.25 Q$ is shown by the vertical dashed line). It is interesting to observe that the TMD factorization formula works unexpectedly well outside of this region.

\paragraph{Normalization issue}

The main problem of presented analysis is the significant difference in the common value (normalization) between the theory prediction and E615 measurement. For a deeper understanding of this issue, it is instructive to perform the decomposition of $\chi^2$ values as
\begin{eqnarray}\label{ex:chi=1+2}
\chi^2=\chi_D^2+\chi_\lambda^2,
\end{eqnarray}
where $\chi_D^2$($\chi_\lambda^2$) represents the uncorrelated(correlated) part of $\chi^2$. Loosely speaking, the value of $\chi_D^2$($\chi_\lambda^2$) demonstrates the agreement in the shape (normalization) between the theory and the data. The decomposition (\ref{ex:chi=1+2}) is done with the help of nuisance parameters \cite{Ball:2008by,Ball:2012wy}. As a by-product, this method allows determining the value of so-called ``systematics shifts'' $d_i$ that are the deviation between the theory and the data due to the normalization only. The results of the nuisance-parameters-decomposition, as well as, average values of $d_i$ are presented in fig.\ref{fig:E537-dX},\ref{fig:E615-dX},\ref{fig:E537-dQ},\ref{fig:E615-dQ} for each bin for E615 and common for E537. 

\begin{figure}[t]
\begin{center}
\includegraphics[width=0.4\textwidth]{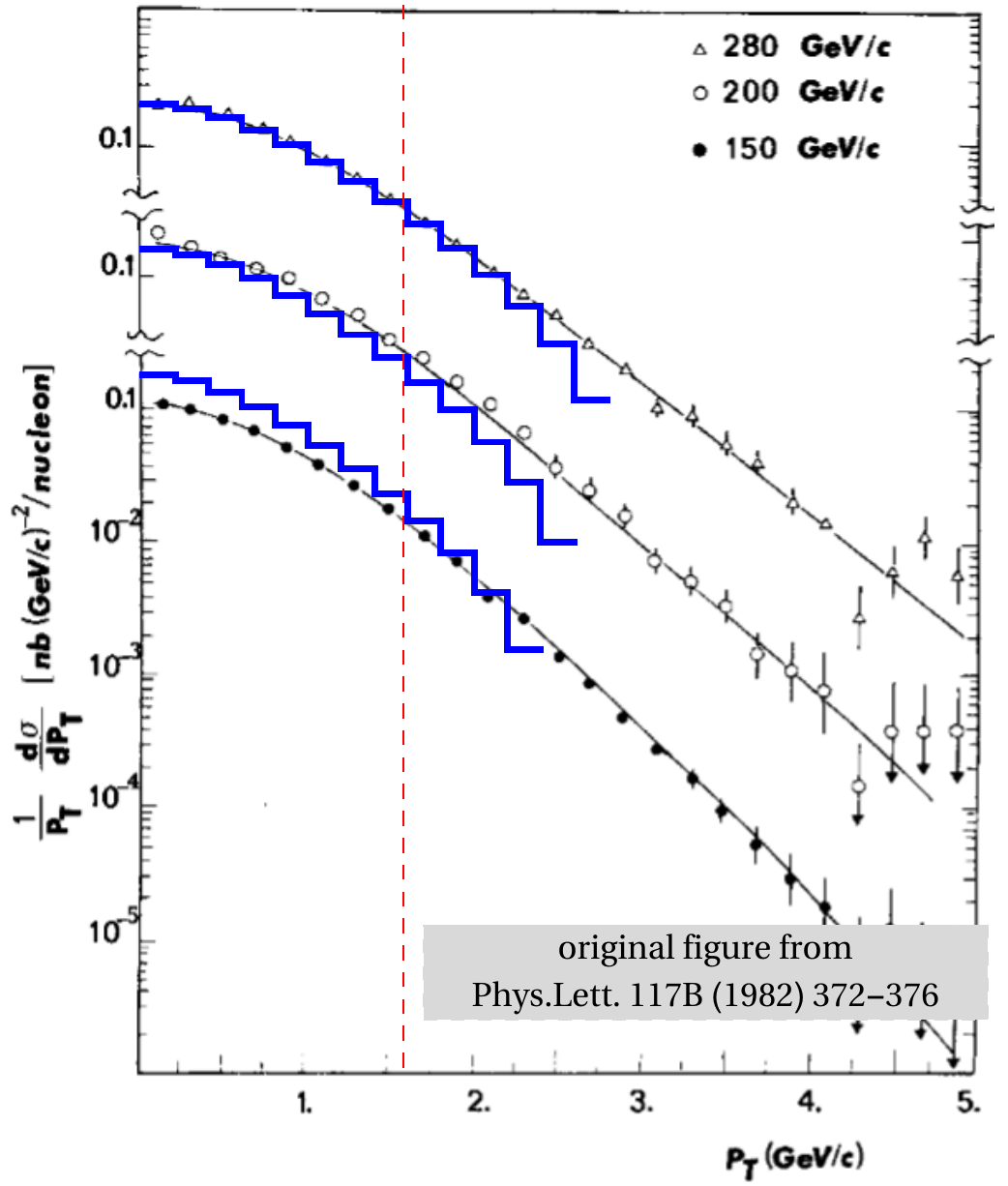}
~~~
\includegraphics[width=0.4\textwidth]{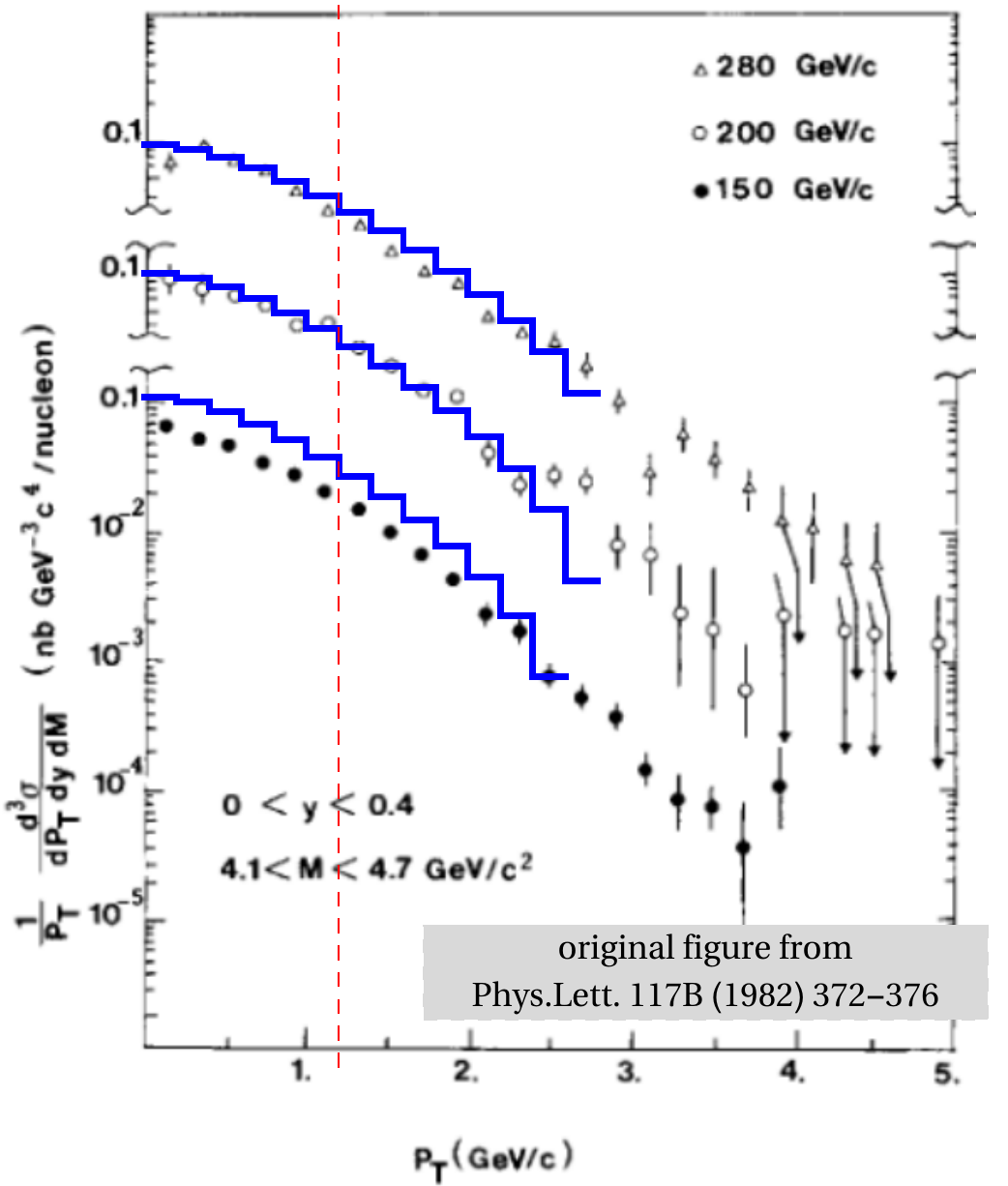}
\end{center}
\caption{\label{fig:NA3} Comparison of the theory prediction (solid line) to NA3 measurement. The theory prediction is plot on the top of figs. 1 and 2 by ref.\cite{Badier:1982zb}. The vertical dashed line shows the estimation of the boundary for TMD factorization approach.}
\end{figure}

The decomposition of $\chi^2$ for the selected data is
\begin{eqnarray}
\chi^2/N_p=0.67+0.77=1.44.
\end{eqnarray}
The value $\chi_\lambda^2/N_p=0.77$ is huge, accounting 16\% systematic uncertainty. Indeed, figures \ref{fig:E615-dX} and \ref{fig:E615-dQ} clearly demonstrates that the theory prediction is systematically below the data. For the first bins (the lowest $x_F$ and $Q$) the difference is practically factor 2. The comparison to E537 (fig.\ref{fig:E537-dX} and \ref{fig:E537-dQ} does not show such a significant problem, but the quality of E537 measurement is much worse. The visual comparison to NA3 measurement (fig.\ref{fig:NA3}) also does not show any normalization problem. Neglecting the normalization part of the $\chi^2$ the agreement between the data and the theory is almost perfect, which is also clear from comparison of dashed lines to data-points in fig.\ref{fig:E537-dX},\ref{fig:E615-dX},\ref{fig:E537-dQ},\ref{fig:E615-dQ}.

The analogous problem with the description of the transverse momentum spectrum for the Drell-Yan process has been recently discussed in ref.\cite{Bacchetta:2019tcu}. The authors of ref.\cite{Bacchetta:2019tcu} have observed that the data-points measured in the fixed-target experiments are significantly (2-3 times) above the theory expectations. The data analyzed in ref.\cite{Bacchetta:2019tcu} belong to the same kinematic domain as the data discussed here. The comparison has been done in the regime $q_T\sim Q$ where the collinear factorization is well established. The authors have tested several ways to improve the theory predictions (threshold resummation, $k_T$-smearing) but were not able to resolve the problem. In the TMD regime the same effect has been observed in \cite{Bertone:2019nxa} (for the same experiments that are considered in \cite{Bacchetta:2019tcu}), namely, about 40\% deficit in the normalization that decreases with the increase of energy (see table 3 in  \cite{Bertone:2019nxa}). Note, both analyses \cite{Bacchetta:2019tcu} and \cite{Bertone:2019nxa} have not a problem with the description of PHENIX data \cite{Aidala:2018ajl} that have a similar range of $Q$ but measured in the collider regime. A similar problem was also observed in semi-inclusive deep-inelastic scattering (SIDIS) \cite{Gonzalez-Hernandez:2018ipj}.

Previously, E615 measurement have been analyzed in the framework of TMD factorization in refs.\cite{Wang:2017zym} and \cite{Ceccopieri:2018nop}. In these articles, authors do not observe any problems with the normalization. However, in both cases, functions used to fit the non-perturbatibatve parts include parameters that significantly influence the normalization. Therefore, it is possible that the normalization issue discussed here, was absorbed into model parameters in refs.\cite{Wang:2017zym,Ceccopieri:2018nop}. 

The present situation could appear because of the problem with the theory. Let me list possible flaws of the current consideration
\begin{itemize}
\item Nuclear effects. The nuclear effects are, for sure, presented in current measurements and goes beyond iso-spin modification (\ref{th:iso-spin-rotation}). Generally, an extra factor $R_i^A(x)$ for PDF should be added, see e.g.\cite{Eskola:2016oht}. Typically, at $x\in(0.1,0.9)$ this factor provides $\sim 10\%$ modification \cite{Eskola:2016oht}, which cannot compensate the gap between the theory and the data. Moreover, this effect should be much smaller for $x$-integrated bins (fig.\ref{fig:E615-dQ}) due to the oscillation of $R_i^A(x)$ between anti-shadowing and EMC regimes. 
\item Effects of PDF. The collinear PDFs are poorly known at large-$x$, and values of PDF significantly differ between different sets. In particular, the difference between PDF values at large-x completely resolves the normalization issue (of the order of $5\%$) with LHCb Z-boson spectrum in \cite{Bertone:2019nxa,TOBE}. I have checked that in the present kinematics the usage of different PDF sets could produce up to $20\%$ difference at a point. Even so, it mainly affects the shape of the cross-section, whereas the normalization is affected only by $2-3$\%. Note, that the pion PDF were extracted mainly from the integrated over $q_T$ measurement by E615 \cite{Barry:2018ort}, and in the present analysis TMDPDF accurately (at NNLO) matches collinear PDF.
\item Threshold contributions. The large-$x$ effects must be incorporated into the matching coefficient in (\ref{th:TMDPDF}). To my opinion, the ignorance of threshold effects leads to the disagreement in the shape of cross-section for bins with $x_F>0.7$ (fig.\ref{fig:E537-dX},\ref{fig:E615-dX}). However, the effect of threshold resummation should be negligible at $x\sim 0.2$ and $Q\sim 4-5$GeV, where the most significant deviation takes place. Also in ref.\cite{Bacchetta:2019tcu} a more accurate analysis has been performed, and it has been shown that the threshold resummation does not solve the problem
\item Power corrections. The TMD factorization theorem violates QED Ward identities and Lorentz invariance (it is typical for factorization theorems with several scales, see e.g., discussion in \cite{Braun:2011dg}). To restore it, one needs to account power corrections, which could be large. Nowadays, there are no systematic studies of power corrections to TMD factorization, and their size is unknown. Nonetheless, these corrections must vanish at $q_T/Q\to 0$, and so, their presence would be indicated in the deformation of the shape of cross-section, what is not observed.
\item Wrong shape for non-perturbative corrections. It could happen that the suggested ansatz for non-perturbative parts of TMD evolution (\ref{th:DNP}) and TMDPDF (\ref{th:TMDPDF}) is essentially wrong, and confines the cross-section in improper domain. However, it looks very implausible because it agrees with known theory constraints, and nicely describe the proton-proton measurements \cite{Bertone:2019nxa}.
\item Resonance effects. The most problematic bins are the lower-$Q$ bins. It could imply that the observed deficit in the normalization is produced by the interference of $\gamma^*$ with $J/\psi$, $\psi'$ resonances and their excitations that are located in the region $Q\sim 3-4$GeV. However, the post-resonance contamination typically looks exactly opposite, as an excess of the theory over the data.
\end{itemize}
In total, it is hard to imagine that any of these points (except resonance contamination) could change the value of cross-section normalization more than 5-10\%. Unless the TMD factorization formula has a deep and systematic problem.

Thus, I should conclude that probably the differential in $q_T$ data by E615 have an incorrect normalization. There are some details that further point to this possibility. First, there is a very good agreement in the shape of cross-sections. Second, the normalization issue is greater at smaller-$Q$ and practically disappears at $Q\sim 9$GeV (the same with $x_F$-differential bins since $x_F\sim Q/\sqrt{s}$). It could indicate the bad estimation of the background in the close-to-resonance region by E615 collaboration. Additionally, the traces of abnormal behavior in $x_F$ (for $q_T$-spectrum) are already seen in the publication of E615 \cite{Conway:1989fs}. It was observed that $q_T$-spectrum after subtraction of normalization has an extreme dependence on $x_F$ (see sec.V.B, and appendix A, in ref.\cite{Conway:1989fs}), which could not be explained within the perturbative QCD. Finally, the comparison to E537 and NA3 experiments has not a problem with normalization, although data-quality is significantly worse.

\section{Conclusion}

\begin{figure}[t]
\begin{center}
\includegraphics[width=0.4\textwidth]{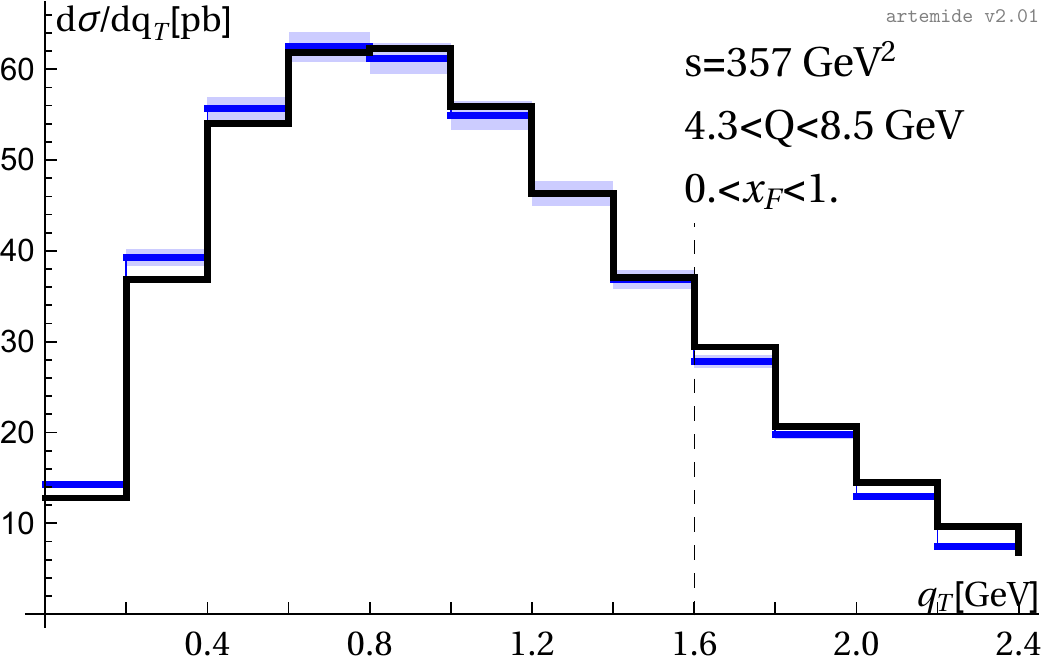}
~~
\includegraphics[width=0.4\textwidth]{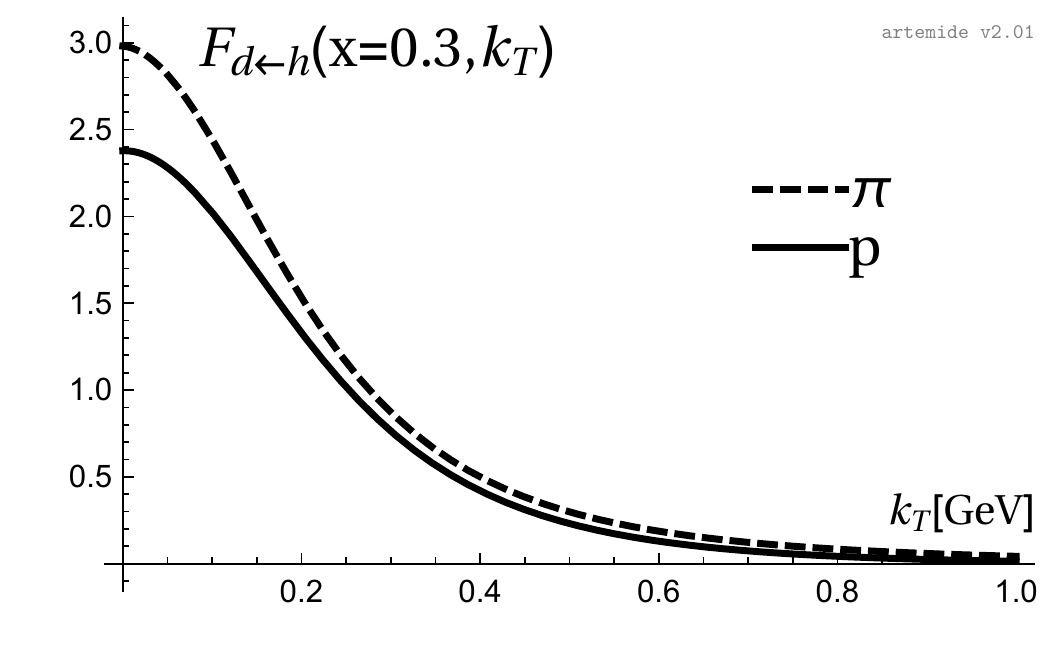}
\end{center}
\caption{\label{fig:COMPASS} (left) Comparison of theory prediction to the preliminary results of COMPASS \cite{Aghasyan:2017jop}. The experimental values are normalized to the theory. Vertical line shows approximate boundary of TMD factorization approach. (right) Comparison of unpolarized TMDPDF of d-quark in pion and proton at $x=0.3$.}
\end{figure}

In the present work, the pion-induced Drell-Yan process has been studied, with the main aim to extract the values of pion unpolarized transverse momentum dependent parton distribution function (TMDPDF). The analysis is made in the TMD factorization framework with $\zeta$-prescription \cite{Scimemi:2018xaf} and compete next-to-next-to-leading (NNLO) perturbative input. To extract the values of pion TMDPDF, the measurements of E615 experiment have been used. I have used the differential in $x_F$ data for better sensitivity to $x$-dependence of TMDPDF. The measurement of E615 differential in Q and measurements by E537 and NA3 were used for the cross-check of the fit. The resulting pion TMDPDFs are available as a part of \texttt{artemide} (model \texttt{Vpion19}) -- the program package for TMD phenomenology \cite{web}.

During the fit procedure, I have faced the problem of systematic disagreement in the normalization between data and the theory. The measurements with low-$Q$ and, correspondingly low-$x_F$, are significantly higher (up to two times for $Q\sim 4-5$GeV) than the prediction. Simultaneously the shape of cross-sections is in an excellent agreement. The size of discrepancy in the normalization decreases with the increase of the $Q$. In the last part of sec.\ref{sec:fit}, I provide a discussion on possible sources of normalization disagreement and conclude that I do not see any possibility to obtain such a significant factor within the modern TMD factorization framework. There is a possibility that the observed normalization problem has an experimental origin. The comparison of the theory with E537 and NA3 has not such a problem, but the both experiments have much worse precision, and could not seriously compete with E615. A similar problem has been recently observed in the TMD spectrum of proton-nucleus Drell-Yan process in \cite{Bacchetta:2019tcu}.

Within the nearest future, the COMPASS collaboration will repeat the analysis of the pion-induced Drell-Yan process in the similar kinematics regime. The announcement of this measurement is presented in \cite{Aghasyan:2017jop}. In fig.\ref{fig:COMPASS}(left) the comparison of the preliminary COMPASS data to the prediction made with \texttt{Vpion19} is shown. Hopefully, the COMPASS measurement will resolve the problem with the normalization of E615 experiment.

A particularly engaging point to study pion TMDPDF is its comparison to proton TMDPDF since the confined motion of partons in mesons and baryons could be fundamentally different. However, any principal difference is not observed (at moderate $x$), see fig.\ref{fig:COMPASS}(right). At high-$x$ distributions looks different, but no conclusion can be done since high-$x$ region is not well controlled both experimentally and theoretically. Definitely, the future measurements of TMD cross-section for pion-induced Drell-Yan process will shed light to this side of parton dynamics.

\acknowledgments I thank Wen-Chen Chang for the correspondence that initiated this work, and for critical remarks and suggestions.

\appendix

\section{Special null-evolution line at large $b$}
\label{app:app1}

The concept of the special null-evolution line plays the central role in $\zeta$-prescription. The $\zeta$-prescription, the double evolution and properties of TMD evolution have been elaborated in ref.\cite{Scimemi:2018xaf}, where I refer for further details. In this appendix, I derive the (perturbative) expression for the special null-evolution line that exactly incorporates non-perturbative corrections.

A null-evolution line is defined as an equipotential line for the 2-dimensional field of anomalous dimensions $\mathbf{E}=(\gamma_F(\mu,\zeta)/2,-\mathcal{D}(\mu,b))$ in the plane $(\mu,\zeta)$. The anomalous dimension $\gamma_F$ is the ultraviolet anomalous dimension of TMD operator. It has the following form
\begin{eqnarray}
\gamma_F(\mu,\zeta)=\Gamma_{\text{cusp}}(\mu)\ln\(\frac{\mu^2}{\zeta}\)-\gamma_V(\mu),
\end{eqnarray}
where $\Gamma_{\text{cusp}}$ is the cusp-anomalous dimension, and $\gamma_V$ anomalous dimension of the vector form-factor. The rapidity anomalous dimension $\mathcal{D}(\mu,b)$ is generally non-perturbative function, which can be computed perturbatively only at small-b, see e.g.\cite{Vladimirov:2016dll,Echevarria:2015byo} for NNLO and N$^3$LO computations. It satisfies the renormalization group equation
\begin{eqnarray}
\mu^2 \frac{d \mathcal{D}(\mu,b)}{d\mu^2}=\frac{\Gamma_{\text{cusp}}(\mu)}{2}.
\end{eqnarray}
Due to this expression the field $\mathbf{E}$ is conservative. Parameterizing an equipotential line as $(\mu,\zeta(\mu,b))$, one finds the following equation for $\zeta(\mu,b)$
\begin{eqnarray}\label{th:sp-line}
\Gamma_{\text{cusp}}(\mu)\ln\(\frac{\mu^2}{\zeta(\mu,b)}\)-\gamma_V(\mu)=2\mathcal{D}(\mu,b) \frac{d \ln \zeta(\mu,b)}{d \ln \mu^2}.
\end{eqnarray}
The special null-evolution line is the line that passes thorough the saddle point $(\mu_0,\zeta_0)$ of the evolution field. The saddle point is defined as
\begin{eqnarray}
\mathcal{D}(\mu_0,b)=0,\qquad \gamma_F(\mu_0,\zeta_0)=0.
\end{eqnarray}
Such boundary condition are very important for two reasons. First, there is only one saddle point in the evolution field, and thus, the special null-evolution line is unique. Second, the special null-evolution line is the only null-evolution line, which has finite $\zeta$ at all values of $\mu$ (bigger than $\Lambda_{QCD}$). It follows from the definition of the saddle point, and guaranties the finiteness of perturbative series order-by-order. 

The field $\mathbf{E}$, and consequently the equipotential line $\zeta(\mu)$ and the position of the saddle point $(\mu_0,\zeta_0)$, depends on $b$, which is treated as a free parameter. It causes certain problems in the implementation of the $\zeta$-prescription. The lesser problem is that additional numerical computations are required to determine the position of saddle-point and the values of the line for different non-perturbative models of $\mathcal{D}$. The greater problem is that at larger $b$ the value of $\mu_0$ decreases and at some large value of $b$ (typically $b\sim 3$GeV$^{-1}$) $\mu_0$ is smaller than $\Lambda_{QCD}$.  Due to this behavior, it is impossible to determine the special null-evolution line at large-$b$ numerically. Note, that nonetheless the special null-evolution line is still uniquely defined by the continuation from smaller values of $b$. In ref.\cite{Bertone:2019nxa} the value of the special null-evolution line has been approximated by perturbative expression with $b=f(b)$, which exactly matches true values at $b\to 0$, and starts to significantly deviate from exact values at $b\sim 3-4$GeV$^{-1}$. This deviation has been considered as a part of non-perturbative model for evolution, which somewhat undermine universality of non-perturbative TMD evolution kernel, and adds correlation between non-perturbative parts of TMD evolution kernel and TMDPDFs. Recently, I have found a simple solution for the problem of determination of the special null-evolution line, which is presented here.

The main breakthrough idea is to use the non-perturbative rapidity anomalous dimension as a generalized coordinate instead of the scale $\mu$. It could not be done entirely, since scale $\mu$ also enters QCD coupling constant in anomalous dimensions $\Gamma_{\text{cusp}}$ and $\gamma_V$. For values of $\mu$ large-enough the value of $a_s(\mu)$ is small, and thus the solution could be evaluated order-by-order in $a_s(\mu)$. Important, that the non-perturbative dependence is exactly accounted in such approach. The equation (\ref{th:sp-line}) can be rewritten
\begin{eqnarray}\label{app:g-}
2\mathcal{D} \(1+\beta(a_s)\frac{\partial g(a_s,\mathcal{D})}{\partial a_s}-\frac{\Gamma_{\text{cusp}}(a_s)}{2}\frac{\partial g(a_s,\mathcal{D})}{\partial \mathcal{D}}\)
-\Gamma_{\text{cusp}}(a_s)g(a_s,\mathcal{D})+\gamma_V(a_s)=0,
\end{eqnarray}
where $g(\mu,b)=\ln(\mu^2/\zeta(\mu,b))$, and $\beta$ is QCD beta-function. In this terms the boundary condition turns into finiteness of function $g$ at $\mathcal{D}=0$. The equation (\ref{app:g-}) can be easily solved order-by-order in $a_s$. Denoting
\begin{eqnarray}\label{app:g}
g(a_s,\mathcal{D})=\frac{1}{a_s}\sum_{n=0}^\infty a_s^n g_n(\mathcal{D}),
\end{eqnarray}
\begin{eqnarray}\nn
\beta(a_s)=\sum_{n=0}^\infty a_s^{n+2}\beta_n,\qquad \Gamma_{\text{cusp}}(a_s)=\sum_{n=0}^\infty a_s^{n+1}\Gamma_n,\qquad \gamma_V(a_s)=\sum_{n=1}^\infty a_s^n \gamma_n.
\end{eqnarray}
I obtain
\begin{eqnarray}\label{app:g0}
g_0&=&\frac{e^{-p}+p-1}{\beta_0 p},
\\
g_1&=&g_0\(\frac{\beta_1}{\beta_0}-\frac{\Gamma_1}{\Gamma_0}\)+\frac{\gamma_1}{\gamma_0}-\frac{\beta_1}{2\beta_0^2}p,
\\\label{app:g2}
g_2&=&g_0\frac{\beta_2\Gamma_0-\beta_1\Gamma_1}{\beta_0\Gamma_0}+\frac{\ch \,p-1}{p}\frac{\beta_0\Gamma_1^2-\beta_0\Gamma_0\Gamma_2+\beta_1\Gamma_0\Gamma_1-\beta_2\Gamma_0}{\beta_0^2\Gamma_0^2}+\frac{e^p-1}{p}\frac{\Gamma_0\gamma_2-\Gamma_1\gamma_1}{\Gamma_0^2},
\end{eqnarray}
where $p=2\beta_0 \mathcal{D}/\Gamma_0$. Let me mention that NNLO term is exponentially grow at large-$\mathcal{D}$ (the N$^3$LO term grows even faster as $e^{2p}$). However, it is not a problem, since i) $g$ enters the logarithm, ii) asymptotic regime takes place at very large values of $b$, iii) altogether such behavior only suppresses high-b tale of the evolution exponent. The expressions (\ref{app:g0}-\ref{app:g2}) provide a very accurate approximation, since $a_s$ is evaluated at $\mu=Q$ and typically $a_s=g^2/(4\pi)^2\sim 10^{-2}$. The most important is that this expression is valid at all values of $b$, even then saddle point is below $\Lambda_{QCD}$.

\bibliographystyle{JHEP} 
\bibliography{TMD_ref}

\end{document}